# Electroluminescence from chirality-sorted (9,7)-semiconducting carbon nanotube devices

Martin H.P. Pfeiffer,[1,2] Ninette Stürzl,[1] Christoph W. Marquardt,[1] Michael Engel,[1] Simone Dehm,[1] Frank Hennrich,[1] Manfred M. Kappes,[1,3,4] Uli Lemmer,[2,4] and Ralph Krupke[1,4,*]

[1]*Institute of Nanotechnology, Karlsruhe Institute of Technology, 76021 Karlsruhe, Germany*
[2]*Lichttechnisches Institut, Karlsruhe Institute of Technology, 76131 Karlsruhe, Germany*
[3]*Institut für Physikalische Chemie, Karlsruhe Institute of Technology, 76131 Karlsruhe, Germany*
[4]*Center for Functional Nanostructures, 76131 Karlsruhe, Germany*
*\*krupke@kit.edu*

**Abstract**: We have measured the electroluminescence and photoluminescence of (9,7)-semiconducting carbon nanotube devices and demonstrate that the electroluminescence wavelength is determined by the nanotube's chiral index (n,m). The devices were fabricated on $Si_3N_4$-membranes by dielectrophoretic assembly of tubes from monochiral dispersion. Electrically driven (9,7)-devices exhibit a single Lorentzian-shaped emission peak at 825 nm in the visible part of the spectrum. The emission could be assigned to the excitonic E22 interband-transition by comparison of the electroluminescence spectra with corresponding photoluminescence excitation maps. We show a linear dependence of the EL peak width on the electrical current, and provide evidence for the inertness of $Si_3N_4$ surfaces with respect to the nanotubes optical properties.

**1. Introduction**

Semiconducting single-walled carbon nanotubes (sem-SWNT) are of interest for optoelectronic applications because of their direct bandgap that can be used for the generation and the detection of light.[1] Their excited states are excitonic in nature with binding energies on the order of 0.5eV,[2,3] which can be generated optically and electrically. Excitons can dissociate at interfaces and generate photocurrents,[4] recombine radiatively and generate light, or decay non-radiatively and produce heat.[5,6]

The photoluminescence (PL) spectrum is uniquely determined by the nanotube diameter and its chiral angle. PL excitation maps probing the first (E11) and second (E22) excited state are thus used to identify the chiral index (n,m) of sem-SWNTs.[7] Photocurrent and electroluminescence spectra show characteristic peaks that were associated with the generation of excitons.[8,9,10] Recently, Adam has measured the electroluminescence (EL) of carbon nanotube networks and small bundles prepared from unsorted nanotubes and showed that the emission wavelength in such samples is determined by the largest diameter sem-SWNTs within a sample.[11]

Here we take advantage of recent progress in fabricating single-chirality nanotube devices and present PL and EL measurements on (9,7) sem-SWNTs. The devices were fabricated on $Si_3N_4$-membranes by dielectrophoretic deposition from monochiral (9,7) dispersions.[12] The EL spectra show a single Lorentzian-shaped emission line in the visible part of the spectrum which - by comparison with PL data – could unambiguously be assigned to the E22 transition of (9,7) sem-SWNTs. We discuss the EL emission pattern, the evolution of the EL spectra with driving current and relate the EL peak width to the PL spectra. We also comment on the difference of (9,7) PL-spectra measured on $SiO_2$ and $Si_3N_4$ substrates.

## 2. Experimental

The (n,m)-nanotube devices have been prepared by low-frequency dielectrophoresis from single-chirality nanotube dispersions. The source-drain electrodes of Pd/Ti were fabricated by standard electron-beam lithography on 50 nm thin $Si_3N_4$ membranes of 50 μm x 50 μm lateral size (Fig. 1a). The membranes are supported by a silicon substrate frame and the structures were mounted and wire bonded to a ceramic package. The electrode gap size has been set to 500 nm. The conditions for dielectrophoresis and the preparation of (9,7) / poly(9,9-di-n-octylfluorenyl-2,7-diyl) (PFO) / toluene solution have been reported elsewhere.[12] Typically about 5 CNTs have been deposited per electrode pair as characterized by scanning electron microscopy (SEM).

The EL of the devices has been measured under vacuum at room temperature in an optical cryostat coupled to a microscope with spectrometer and silicon charge-coupled-device (CCD) camera, which is sensitive in the wavelength range 200-1050 nm.[13] In our setup the EL signal can be recorded in an imaging and spectroscopy mode, selecting a mirror or a grating, respectively. The devices were operated under current bias with integration times for EL spectra of up to 30 min. All EL spectra have been corrected for the relative spectral response of the optical setup[13]. The PL of the device has been measured with a confocal near-infrared microscope coupled to an InGaAs detector array,[14] and PL excitation maps from devices have been recorded by positioning a tunable laser between the source-drain electrodes and measuring the PL for a range of excitation energie.

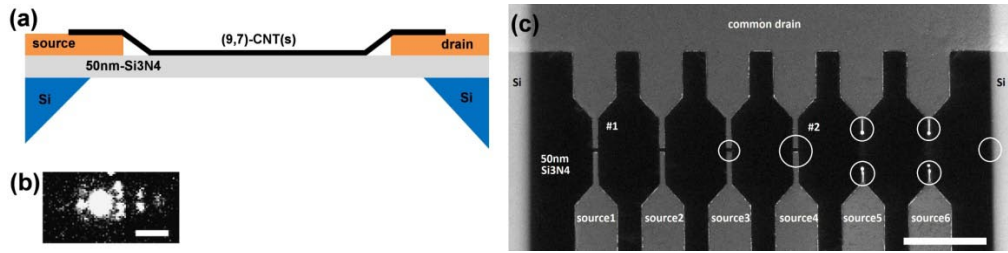

Fig. 1, (a) Schematic cross section of the (9,7)-nanotube device. (b) Top view image of the light emission pattern. The bright emission spot origins from the biased device #2, the weaker spots are caused by scattering of the emitted light at neighboring structures. (c) Corresponding scanning electron micrograph of the device array. The biased device #2 is indicated by a large circle, the structures that cause the light scattering observed in (b) are indicated by small circles. (scale bar length = 10 μm)

## 3. Results

Figure 1c is a scanning electron micrograph (SEM) of a device array with one common drain electrode and independent source electrodes. The thin $Si_3N_4$ membrane appears in the SEM as a black region with virtually zero background intensity due to the very low rate of secondary electrons.[15] Devices show the expected highly non-linear current-voltage (*I-V*) characteristics, as shown in Fig. 3b.

Figure 1b shows the EL of the biased (9,7)-device #2, recorded in the imaging mode. We observe light emanating from several locations although only device #2 has been biased – an effect which is frequently observed on our structures. By comparing Fig. 1b with Fig. 1c, and with EL images recorded under weak external illumination (not shown), we can trace the high-intensity spot back to the position of the biased (9,7)-device #2 (marked by a large circle). The additional emission spots correlate with the positions of the neighboring metal electrodes and with the edge of the Si frame (all marked by small circles): The high-intensity spot is caused by the EL of the biased device, whereas the additional spots are originating from EL light that has been scattered at lateral structures in the vicinity of the biased device.

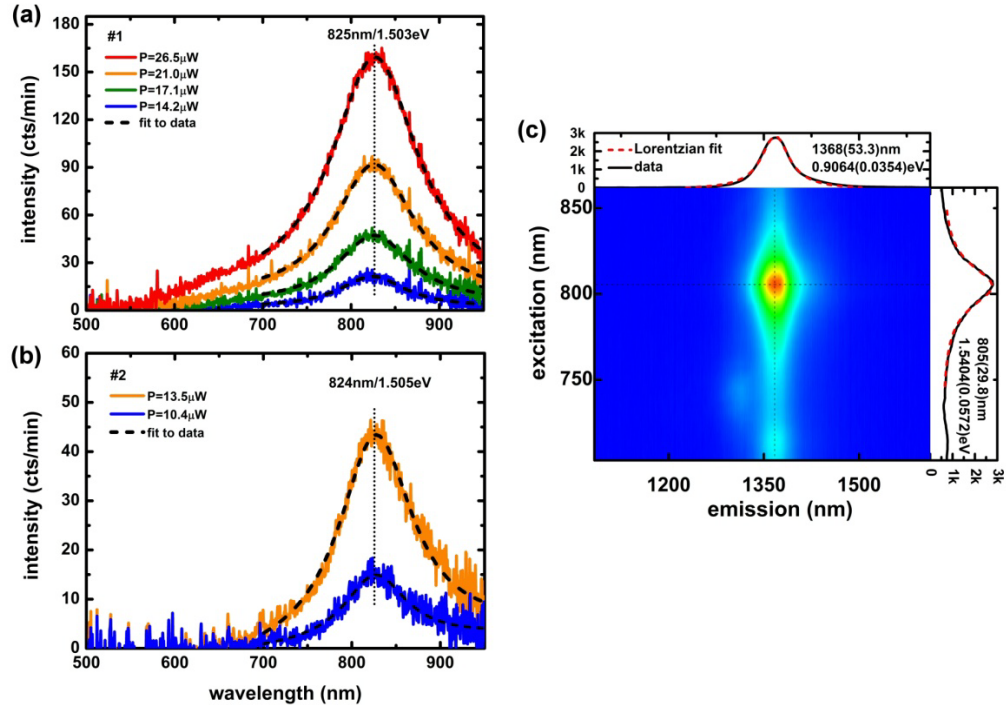

Fig. 2, Evolution of the electroluminescence spectra with driving power of (9,7)-CNT device #1 (a) and #2 (b). (c) Photoluminescence excitation map of a (9,7)-device on $Si_3N_4$. All spectra are fitted to a Lorentzian lineshape.

We discuss now the EL signals that have been further analyzed in the spectroscopy mode. Figure 2a-b shows the evolution of the spectra of two devices #1 and #2 with increasing bias. Both spectra are dominated by a single emission peak at 825 nm, as expected due to the monochirality of the emitter. Both emission peaks can be fitted with a Lorentzian lineshape. Fig. 3a shows that the peak intensities are increasing with the current bias beyond a threshold current of 1-1.5 μA. This behavior is similar to the EL characteristics of CNT p-n junctions and shows that the rate of radiative recombination scales with the current.[16,17] We assume that the mechanism that causes EL here is impact ionization, because of the large applied voltage and the observed threshold for light emission.[10] We note that the peak positions remain constant and that the spectra are stable for hours within the indicated current range.

To assign the EL peak to a particular optical transition we have measured the PL excitation map of a (9,7)-device on the $Si_3N_4$-membrane. Figure 2c shows the maximum intensity of the emission versus excitation wavelength at 1368 nm and 805 nm, corresponding to the excitonic E11 and E22 K-point interband transitions, respectively.[18] The horizontal and vertical cuts through the peak maximum could again be fitted nicely to a Lorentzian lineshape. Similar data has been obtained for (9,7)-devices on thermally oxidized Si substrates albeit with E11 and E22 slightly shifted to 1385 nm and 802 nm, respectively.[12] Interestingly we do not observe a phonon-mediated dark exciton peak (1540 nm) which is very intense for (9,7)-tubes on $SiO_2$ surfaces,[12] demonstrating the strong influence of the silicon surface (likely OH-groups) on the nanotube (optical) properties.

We can now correlate the PL spectra with the EL spectra and assign the EL peak at ~825 nm to the E22 transition of a (9,7)-tube. The EL peak is red-shifted by ~20 nm (~35 meV) of the PL peak, which could be related to electrode-induced screening.[19] We note however that the EL peak position does not depend on the applied bias. A more quantitative analysis of the Lorentzian fits to the EL and PL spectra also shows that the EL peaks are significantly broadened as compared to the PL spectrum. Moreover we observe that the energy width of the EL peaks depends linearly on the source-drain current, and extrapolates for the two samples to a zero-current value that resembles the width of the PL excitation profile (Fig. 3a). It makes sense to correlate the extrapolated zero-bias EL peak width to the width of the PL excitation profile, since both experiments probe the same E22 transition.

The E22 transition (not the E11 transition) and hence the PL excitation profile is entirely determined by lifetime-broadening,[20] and so must be the EL measured in this work – in agreement with the Lorentzian lineshape. Recently a widening of the PL excitation profile at high photon flux has been reported and explained by exciton-exciton-annihilation (EEA) induced lifetime reduction.[21] Our data shows that the EL linewidth depends linearly on the current, reaching values up to 200 meV. It is feasible that EEA increases with current, reduces the exciton lifetime, and thus causes EL lineshape broadening. In this model the EL peak width would be a direct measure of the exciton lifetime.

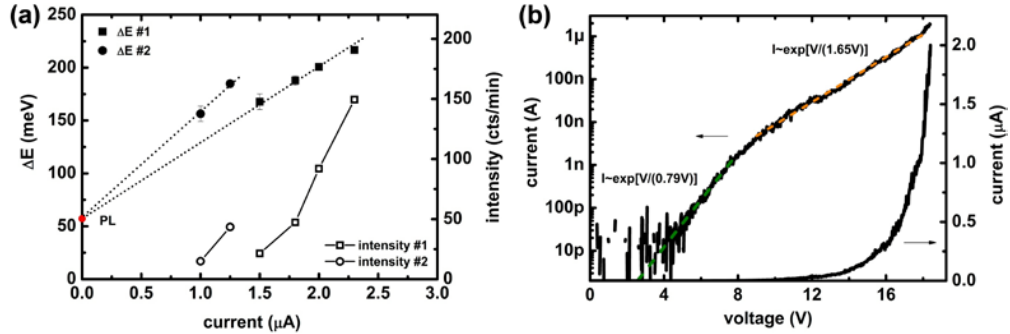

Figure 3: (a) Electroluminescence peak width ΔE and peak intensity versus current for (9,7) CNT device #1 and #2. Marked in red is the width of the photoluminescence excitation profile of Fig.2(c). (b) Current-voltage characteristic of a (9,7) device.

We estimate the efficiency of E22 photon generation, by converting the integrated counts per minute into photons per second, corrected for the absolute sensitivity of our system.[22] For the (9,7)-device #2 we count $3 \cdot 10^9$ photons/sec at 2.3 μA current-bias. This translates into an efficiency of $2 \cdot 10^{-4}$ E22-photons being generated per charge. Taking into account that the integrated EL-intensity ratio of E11/E22 is ~10,[23] the total efficiency is ~$10^{-3}$ which is among the highest values reported for EL through impact ionization.[23] Beneficial for the device performance could be the combination of PFO-coated nanotubes with $Si_3N_4$ substrate. The PL data shows that $Si_3N_4$ has no influence on the optical properties. At the same time is the PFO a strong tunneling barrier,[12] reducing the current at larger voltages and minimizing heat dissipation in the nanotube. Interestingly the current-voltage characteristic (Fig. 3b) shows two exponential slopes with 0.79 V and 1.65 V, which could be assigned to the E11 and E22 single-particle excitation gap that is somewhat larger than the optical gap.[1] As a final remark we note that the Silicon-CCD did not allow a direct study of the E11 emission in the near-infrared part of the spectrum.

In summary we have studied in detail the electroluminescence signal from (9,7)-carbon-nanotube devices and assigned the 825 nm peaks to the excitonic E22 K-point interband transition. The electroluminescence signal is thereby of the shortest wavelength

(highest energy) measured so far for semiconducting carbon nanotubes and demonstrates that EL is controlled by the nanotubes molecular structure. Our results also show that for future optoelectronic applications based on few-tubes- or nanotube-film devices it is essential to use single-chirality sorted carbon nanotubes.

## Acknowledgments

RK acknowledges Mathias Steiner (IBM Research, Yorktown Heights, NY) for helpful discussions.